\documentclass[prd,onecolumn,nofootinbib,showpacs,showkeys]{revtex4}
\newcommand\gothfamily{\usefont{U}{ygoth}{m}{n}}
\DeclareTextFontCommand{\textgoth}{\gothfamily}

\begin{document}

\title{F(R) GRAVITY IN PURELY AFFINE FORMULATION}

\author{{\bf Nikodem J. Pop\l awski}}

\affiliation{Department of Physics, Indiana University, Swain Hall West, 727 East Third Street, Bloomington, IN 47405, USA}
\email{nipoplaw@indiana.edu}

\noindent
{\em International Journal of Modern Physics A}\\
Vol. {\bf 23}, No. 12 (2008) 1891--1901\\
\copyright\,World Scientific Publishing Co.
\vspace{0.4in}

\begin{abstract}
The purely affine, metric-affine and purely metric formulation of general relativity are dynamically equivalent and the relation between them is analogous to the Legendre relation between the Lagrangian and Hamiltonian dynamics.
We show that one cannot construct a dynamically equivalent, purely affine Lagrangian from a metric-affine or metric $F(R)$ Lagrangian, nonlinear in the curvature scalar.
Thus the equivalence between the purely affine picture and the two other formulations does not hold for metric-affine and metric theories of gravity with a nonlinear dependence on the curvature, i.e. $F(R)$ gravity does not have a purely affine formulation.
We also show that this equivalence is restored if the metric tensor is conformally transformed from the Jordan to the Einstein frame, in which $F(R)$ gravity turns into general relativity with a scalar field.
This peculiar behavior of general relativity, among relativistic theories of gravitation, with respect to purely affine, metric-affine and purely metric variation could indicate the physicality of the Einstein frame.
On the other hand, it could explain why this theory cannot interpolate among phenomenological behaviors at different scales.
\end{abstract}


\maketitle

\section{Introduction}
\label{secIntro}

In the {\em purely affine} (Einstein-Eddington) formulation of general relativity~\cite{Ein,Edd1,Edd2,Schr2,Kij,Cat}, a Lagrangian density depends on a torsionless affine connection and the symmetric part of the Ricci tensor.
This formulation defines the metric tensor as the derivative of the Lagrangian density with respect to the symmetrized Ricci tensor, obtaining an algebraic relation between these two tensors.
The field equations are derived by varying the total action with respect to the connection, which gives a differential relation between the connection and the metric tensor and thus a differential equation for the metric.
In the {\em metric-affine} (Einstein-Palatini) formulation~\cite{Pal,Lord,FFR}, both the metric tensor and torsionless connection are independent variables (gravitational potentials), and the field equations are derived by varying the action with respect to these quantities.
The corresponding Lagrangian density is linear in the symmetric part of the Ricci tensor.
In the {\em purely metric} (Einstein-Hilbert) formulation~\cite{Hilb1,Hilb2,Hilb3,LL2,MTW,SG}, the metric tensor is a dynamical variable, the affine connection is the Riemannian connection that depends entirely on the metric, and the field equations are derived by varying the action with respect to the metric tensor.
The corresponding Lagrangian density is linear in the Ricci scalar.

All three formulations of general relativity are dynamically equivalent; the relation between the purely affine and metric-affine picture is analogous to the Legendre relation between the Lagrangian and Hamiltonian dynamics~\cite{FK3,FK4}.
A crucial factor in this equivalence is the linearity of metric-affine and purely metric Lagrangians with respect to curvature.
Although to each metric-affine or purely metric Lagrangian for the gravitational field and matter there corresponds the dynamically equivalent purely affine Lagrangian~\cite{Kij,FK3}, the explicit form of such a Lagrangian is known only for few cases: the cosmological constant~\cite{Edd2}, the Klein-Gordon~\cite{Kij}, Maxwell and Proca fields~\cite{FK1}, and barotropic fluids~\cite{KPT,KM,KW}.
This statement can be generalized to theories of gravitation with purely affine Lagrangians that depend on the torsion tensor, full Ricci tensor and tensor of homothetic curvature~\cite{univ,nonsym}.

In this paper we examine the correspondence between metric-affine Lagrangians for the gravitational field that are nonlinear functions of the curvature scalar and the purely affine formulation.
These phenomenological $F(R)$ gravity models are currently of physical interest since they explain inflation and the present acceleration of the universe~\cite{fR1,fR2,fR3,fR4,fR5,fR6,fR7,fR8,fR9,fR10,fR11,fR12,Niko0,fR13,SL,fR14,fR15,fR16,fR17,fR18,SL,fR19,fR20,fR21,fR22,fR23,fR24,fR25,fR26,fR27}.
We show that one cannot construct a purely affine Lagrangian that is dynamically equivalent to a nonlinear (in curvature) metric-affine or metric $F(R)$ Lagrangian unless the metric tensor is conformally transformed from the Jordan frame to the Einstein frame~\cite{EJ1,EJ2,EJ3}.
Thus $F(R)$ gravity, which corresponds to the Hamiltonian formulation of classical mechanics, can be written in the purely affine picture, which is analogous to the Lagrangian formulation, only if we redefine the metric tensor.

\section{Field Equations in Purely Affine Gravity}
\label{secField}

A general purely affine Lagrangian density $\textgoth{L}$ depends on the affine connection $\Gamma^{\,\,\rho}_{\mu\,\nu}$ and the curvature tensor, $R^\rho_{\phantom{\rho}\mu\sigma\nu}=\Gamma^{\,\,\rho}_{\mu\,\nu,\sigma}-\Gamma^{\,\,\rho}_{\mu\,\sigma,\nu}+\Gamma^{\,\,\kappa}_{\mu\,\nu}\Gamma^{\,\,\rho}_{\kappa\,\sigma}-\Gamma^{\,\,\kappa}_{\mu\,\sigma}\Gamma^{\,\,\rho}_{\kappa\,\nu}$.
We assume that the dependence of the Lagrangian on the curvature is restricted to the contracted curvature tensors~\cite{nonsym}: the symmetric $P_{\mu\nu}=R_{(\mu\nu)}$ and antisymmetric $R_{[\mu\nu]}$ part of the Ricci tensor, $R_{\mu\nu}=R^\rho_{\phantom{\rho}\mu\rho\nu}$, and the antisymmetric tensor of homothetic curvature (segmental curvature tensor), $Q_{\mu\nu}=R^\rho_{\phantom{\rho}\rho\mu\nu}=\Gamma^{\,\,\rho}_{\rho\,\nu,\mu}-\Gamma^{\,\,\rho}_{\rho\,\mu,\nu}$, which has the form of a curl~\cite{Schr2,Scho}.
The metric structure associated with a purely affine Lagrangian is obtained using~\cite{Edd1,Edd2,Schr2,Kij,FK3,FK1,FK2,Niko1,Niko2}
\begin{equation}
{\sf g}^{\mu\nu}\equiv-2\kappa\frac{\partial\textgoth{L}}{\partial P_{\mu\nu}},
\label{met1}
\end{equation}
where ${\sf g}^{\mu\nu}$ is the symmetric fundamental tensor density and $\kappa=\frac{8\pi G}{c^4}$.
The symmetric contravariant metric tensor is defined by
\begin{equation}
g^{\mu\nu}\equiv\frac{{\sf g}^{\mu\nu}}{\sqrt{-\mbox{det}{\sf g}^{\rho\sigma}}}.
\label{met2}
\end{equation}
To make this definition meaningful, we have to assume $\mbox{det}({\sf g}^{\mu\nu})\neq0$, which, if $\mbox{det}({\sf g}^{\mu\nu})<0$, also guarantees that the tensor $g^{\mu\nu}$ has the Lorentzian signature $(+,-,-,-)$~\cite{FK2}.
The symmetric covariant metric tensor $g_{\mu\nu}$ is related to the contravariant metric tensor by $g^{\mu\rho}g_{\nu\rho}=\delta^\mu_\nu$.
The tensors $g^{\mu\nu}$ and $g_{\mu\nu}$ are used for raising and lowering indices.
We also define an antisymmetric tensor density:
\begin{equation}
{\sf h}^{\mu\nu}\equiv-2\kappa\frac{\partial\textgoth{L}}{\partial Q_{\mu\nu}},
\label{amet1}
\end{equation}
and the hypermomentum density conjugate to the connection~\cite{Hehl,Smal,HK,HLS}:
\begin{equation}
\Pi_{\phantom{\mu}\rho\phantom{\nu}}^{\mu\phantom{\rho}\nu}\equiv-2\kappa\frac{\partial\textgoth{L}}{\partial \Gamma^{\,\,\rho}_{\mu\,\nu}},
\label{con1}
\end{equation}
which has the same dimension as the connection.
Let us assume that the Lagrangian density $\textgoth{L}$ does not depend on $R_{[\mu\nu]}$.

If we do not restrict the connection $\Gamma^{\,\,\rho}_{\mu\,\nu}$ to be symmetric~\cite{Hehl,Car}, the variation of the Ricci tensor can be transformed into the variation of the connection by means of the Palatini formula~\cite{Schr2}.
The principle of least action $\delta\int d^4x\textgoth{L}(\Gamma^{\,\,\rho}_{\mu\,\nu},P_{\mu\nu},Q_{\mu\nu})=0$ yields~\cite{unif}
\begin{equation}
{\sf g}^{\mu\nu}_{\phantom{\mu\nu},\rho}+\,^\ast\Gamma^{\,\,\mu}_{\sigma\,\rho}{\sf g}^{\sigma\nu}+\,^\ast\Gamma^{\,\,\nu}_{\rho\,\sigma}{\sf g}^{\mu\sigma}-\,^\ast\Gamma^{\,\,\sigma}_{\sigma\,\rho}{\sf g}^{\mu\nu}=\Pi_{\phantom{\mu}\rho\phantom{\nu}}^{\mu\phantom{\rho}\nu}-\frac{1}{3}\Pi_{\phantom{\mu}\sigma\phantom{\sigma}}^{\mu\phantom{\sigma}\sigma}\delta^\nu_\rho+2{\sf h}^{\nu\sigma}_{\phantom{\nu\sigma},\sigma}\delta^\mu_\rho-\frac{2}{3}{\sf h}^{\mu\sigma}_{\phantom{\mu\sigma},\sigma}\delta^\nu_\rho,
\label{field2}
\end{equation}
where $^\ast\Gamma^{\,\,\rho}_{\mu\,\nu}=\Gamma^{\,\,\rho}_{\mu\,\nu}+\frac{2}{3}\delta^\rho_\mu S_\nu$~\cite{Schr2,Schr1}, $S_\mu=S^\nu_{\phantom{\nu}\mu\nu}$ is the torsion vector, $S^\rho_{\phantom{\rho}\mu\nu}=\Gamma^{\,\,\,\,\rho}_{[\mu\,\nu]}$ is the torsion tensor and the semicolon denotes the covariant differentiation with respect to $\Gamma^{\,\,\rho}_{\mu\,\nu}$.
Antisymmetrizing and contracting the indices $\mu$ and $\rho$ in Eq.~(\ref{field2}) gives
\begin{equation}
{\sf h}^{\sigma\nu}_{\phantom{\sigma\nu},\sigma}={\sf j}^\nu,
\label{Max1}
\end{equation}
where
\begin{equation}
{\sf j}^\nu\equiv\frac{1}{8}\Pi_{\phantom{\sigma}\sigma\phantom{\nu}}^{\sigma\phantom{\sigma}\nu}.
\label{Max2}
\end{equation}
Equation~(\ref{Max1}) has the form of the Maxwell equations for the electromagnetic field~\cite{FK2,unif}.

The hypermomentum density $\Pi_{\phantom{\mu}\rho\phantom{\nu}}^{\mu\phantom{\rho}\nu}$ represents the {\em source} for the purely affine field equations~\cite{unif}.
Since the tensor density ${\sf h}^{\mu\nu}$ is antisymmetric, the current vector density ${\sf j}^\mu$ must be conserved: ${\sf j}^\mu_{\phantom{\mu},\mu}=0$, which constrains how the connection $\Gamma^{\,\,\rho}_{\mu\,\nu}$ can enter a purely affine Lagrangian density $\textgoth{L}$: $\Pi_{\phantom{\sigma}\sigma\phantom{\nu},\nu}^{\sigma\phantom{\sigma}\nu}=0$.
If $\textgoth{L}$ depends only on $P_{\mu\nu}$, the field equation~(\ref{Max1}) becomes a stronger, algebraic constraint on how the Lagrangian depends on the connection: $\Pi_{\phantom{\sigma}\sigma\phantom{\nu}}^{\sigma\phantom{\sigma}\nu}=0$.
The dependence of a purely affine Lagrangian on the tensor of homothetic curvature replaces this unphysical constraint with a field equation for ${\sf h}^{\mu\nu}$.
Thus physical Lagrangians that depend explicitly on the affine connection should also depend on either $Q_{\mu\nu}$, which restores the projective invariance of the total action without constraining the connection~\cite{unif}.\footnote{
The dependence of $\textgoth{L}$ on the tensor $R_{[\mu\nu]}$ instead of $Q_{\mu\nu}$ also removes the unphysical constraint on the source density $\Pi_{\phantom{\mu}\rho\phantom{\nu}}^{\mu\phantom{\rho}\nu}$.
However, the resulting field equation does not have the form of the Maxwell equations since $R_{[\mu\nu]}$ is not a curl for a general connection.
}

\section{Equivalence of Purely Affine and Metric-Affine/Purely Metric Formulation}
\label{secEquiv}

If we apply to $\textgoth{L}(\Gamma^{\,\,\rho}_{\mu\,\nu},P_{\mu\nu},Q_{\mu\nu})$ the Legendre transformation with respect to $P_{\mu\nu}$~\cite{Kij,FK3}, defining the Hamiltonian density $\textgoth{H}$:
\begin{equation}
\textgoth{H}=\textgoth{L}-\frac{\partial\textgoth{L}}{\partial P_{\mu\nu}}P_{\mu\nu}=\textgoth{L}+\frac{1}{2\kappa}{\sf g}^{\mu\nu}P_{\mu\nu},
\label{Leg1}
\end{equation}
we find that $\textgoth{H}$ is a function of $\Gamma^{\,\,\rho}_{\mu\,\nu}$, ${\sf g}^{\mu\nu}$ and $Q_{\mu\nu}$.
The action variation with respect to ${\sf g}^{\mu\nu}$ yields the first Hamilton equation~\cite{Kij,FK3,unif}:
\begin{equation}
P_{\mu\nu}=2\kappa\frac{\partial\textgoth{H}}{\partial {\sf g}^{\mu\nu}}.
\label{Ham1}
\end{equation}
The variations with respect to $P_{\mu\nu}$ and $Q_{\mu\nu}$ can be transformed to the variation with respect to $\Gamma^{\,\,\rho}_{\mu\,\nu}$ by means of the Palatini formula and the variation of a curl, respectively, giving the second Hamilton equation equivalent to the field equations~(\ref{field2}).

The analogous transformation in classical mechanics goes from a {\em Lagrangian} $L(q^i,\dot{q}^i)$ to a {\em Hamiltonian} $H(q^i,p^i)=p^j\dot{q}^j-L(q^i,\dot{q}^i)$ (or, more precisely, a Routhian since not all the variables are subject to a Legendre transformation~\cite{LL1}) with $p^i=\frac{\partial{L}}{\partial\dot{q}^i}$, where the tensor $P_{\mu\nu}$ corresponds to generalized velocities $\dot{q}^i$ and the density ${\sf g}^{\mu\nu}$ to canonical momenta $p^i$~\cite{Kij,FK3}.
Accordingly, the affine connection plays the role of the configuration $q^i$ and the source density $\Pi_{\phantom{\mu}\rho\phantom{\nu}}^{\mu\phantom{\rho}\nu}$ corresponds to generalized forces $f^i=\frac{\partial{L}}{\partial q^i}$~\cite{Kij}.
The field equations~(\ref{field2}) correspond to the Lagrange equations which result from Hamilton's principle $\delta\int L(q^i,\dot{q}^i)dt=0$ for arbitrary variations $\delta q^i$ vanishing at the boundaries of the configuration, while the Hamilton equations result from the same principle written as $\delta\int(p^j\dot{q}^j-H(q^i,p^i))dt=0$ for arbitrary variations $\delta q^i$ and $\delta p^i$~\cite{LL1}.
The field equations~(\ref{field2}) correspond to the second Hamilton equation, $\dot{p}^i=-\frac{\partial H}{\partial q^i}$, and Eq.~(\ref{Ham1}) to the first Hamilton equation, $\dot{q}^i=\frac{\partial H}{\partial p^i}$.

If we identify $\textgoth{H}$ with the Lagrangian density for matter $\mathcal{L}_{MA}$ in the metric-affine formulation of gravitation~\cite{FK3}\footnote{
This identification means that the purely affine theory based on the Lagrangian $\textgoth{L}$ and the metric-affine theory based on the matter Lagrangian $\mathcal{L}_{MA}$ equal to $\textgoth{H}$ have the same solutions to the corresponding field equations.
}
then Eq.~(\ref{Ham1}) has the form of the Einstein equations of general relativity, $P_{\mu\nu}-\frac{1}{2}Pg_{\mu\nu}=\kappa T_{\mu\nu}$, where $P=P_{\mu\nu}g^{\mu\nu}$ and the symmetric energy-momentum tensor $T_{\mu\nu}$ is defined by the variational relation: $2\kappa\delta\mathcal{L}_{MA}=T_{\mu\nu}\delta{\sf g}^{\mu\nu}$.
From Eq.~(\ref{Leg1}) it follows that $-\frac{1}{2\kappa}P\sqrt{-g}$, where $g=\mbox{det}g_{\mu\nu}$, is the metric-affine Lagrangian density for the gravitational field $\mathcal{L}_g$, in agreement with the general-relativistic form.
The transition from the affine to the metric-affine formalism shows that the gravitational Lagrangian density $\mathcal{L}_g$ is a {\em Legendre term} corresponding to $p^j\dot{q}^j$ in classical mechanics~\cite{Kij}.
Therefore the purely affine and metric-affine formulation of gravitation are dynamically equivalent if $\textgoth{L}$ depends on the affine connection, the symmetric part of the Ricci tensor~\cite{FK3} and the tensor of homothetic curvature~\cite{nonsym}.

Substituting Eq.~(\ref{Max2}) to~(\ref{field2}) gives a linear algebraic equation for $^\ast\Gamma^{\,\,\,\,\rho}_{\mu\,\nu}$ as a function of the metric tensor, its first derivatives and the density $\Pi_{\phantom{\mu}\rho\phantom{\nu}}^{\mu\phantom{\rho}\nu}$.
The general solution of this equation is given in Refs.~\cite{unif} and~\cite{PO}.
If there are no sources, $\Pi_{\phantom{\mu}\rho\phantom{\nu}}^{\mu\phantom{\rho}\nu}=0$, the connection $\Gamma^{\,\,\rho}_{\mu\,\nu}$ depends only on the metric tensor $g_{\mu\nu}$ representing a free gravitational field and the torsion vector $S_\mu$ corresponding to the vectorial degree of freedom associated with the projective invariance.

The purely metric formulation of gravitation is dynamically equivalent to the purely affine and metric-affine formulation, which can be shown by applying to $\textgoth{H}(\Gamma^{\,\,\rho}_{\mu\,\nu},{\sf g}^{\mu\nu},Q_{\mu\nu})$ the Legendre transformation with respect to $\Gamma^{\,\,\rho}_{\mu\,\nu}$~\cite{FK3}.
This transformation defines the Lagrangian density in the momentum space $\textgoth{K}$:
\begin{equation}
\textgoth{K}=\textgoth{H}-\frac{\partial\textgoth{H}}{\partial\Gamma^{\,\,\rho}_{\mu\,\nu}}\Gamma^{\,\,\rho}_{\mu\,\nu}=\textgoth{H}+\frac{1}{2\kappa}\Pi_{\phantom{\mu}\rho\phantom{\nu}}^{\mu\phantom{\rho}\nu}\Gamma^{\,\,\rho}_{\mu\,\nu},
\label{Leg2}
\end{equation}
which is a function of ${\sf g}^{\mu\nu}$, $\Pi_{\phantom{\mu}\rho\phantom{\nu}}^{\mu\phantom{\rho}\nu}$ and $Q_{\mu\nu}$.
The action variation with respect to ${\sf g}^{\mu\nu}$ yields the Einstein equations:
\begin{equation}
P_{\mu\nu}=2\kappa\frac{\partial\textgoth{K}}{\partial{\sf g}^{\mu\nu}}.
\label{Ham2}
\end{equation}
The variations with respect to $P_{\mu\nu}$ and $Q_{\mu\nu}$ can be transformed to the variation with respect to $\Gamma^{\,\,\rho}_{\mu\,\nu}$ by means of the Palatini formula and the variation of a curl, respectively, giving the field equations~(\ref{field2}).
Finally, the variation with respect to $\Pi_{\phantom{\mu}\rho\phantom{\nu}}^{\mu\phantom{\rho}\nu}$ gives
\begin{equation}
\Gamma^{\,\,\rho}_{\mu\,\nu}=2\kappa\frac{\partial\textgoth{K}}{\partial\Pi_{\phantom{\mu}\rho\phantom{\nu}}^{\mu\phantom{\rho}\nu}},
\label{Ham3}
\end{equation}
in accordance with Eq.~(\ref{Leg2}).

The analogous transformation in classical mechanics goes from a Hamiltonian $H(q^i,p^i)$ to a momentum Lagrangian $K(p^i,\dot{p}^i)=-f^j q^j-H(q^i,p^i)$.
The equations of motion result from Hamilton's principle written as $\delta\int(p^j\dot{q}^j+f^j q^j+K(p^i,\dot{p}^i))dt=0$.
The quantity $K$ is a Lagrangian with respect to $p^i$ because $p^j\dot{q}^j+f^j q^j$ is a total time derivative and does not affect the action variation.

If we define
\begin{equation}
C_\rho^{\phantom{\rho}\mu\nu}=\Pi_{\phantom{(\mu}\rho\phantom{\nu)}}^{(\mu\phantom{\rho}\nu)}-\frac{1}{3}\delta^{(\mu}_\rho\Pi_{\phantom{\nu)}\sigma\phantom{\sigma}}^{\nu)\phantom{\sigma}\sigma}-\frac{1}{6}\Pi_{\phantom{\sigma}\sigma\phantom{(\mu}}^{\sigma\phantom{\sigma}(\mu}\delta^{\nu)}_\rho-\,^\ast\Gamma^{\,\,\,\,\mu}_{(\sigma\,\rho)}{\sf g}^{\sigma\nu}-\,^\ast\Gamma^{\,\,\,\,\nu}_{(\rho\,\sigma)}{\sf g}^{\mu\sigma}+\,^\ast\Gamma^{\,\,\,\,\sigma}_{(\sigma\,\rho)}{\sf g}^{\mu\nu},
\end{equation}
where the connection $\Gamma^{\,\,\rho}_{\mu\,\nu}$ depends on the source density $\Pi_{\phantom{\mu}\rho\phantom{\nu}}^{\mu\phantom{\rho}\nu}$ via Eq.~(\ref{con1}) or~(\ref{Ham3}), then the field equations~(\ref{field2}) and~(\ref{Max2}) can be written as ${\sf g}^{\mu\nu}_{\phantom{\mu\nu},\rho}=C_\rho^{\phantom{\rho}\mu\nu}$.
Accordingly, $\Pi_{\phantom{\mu}\rho\phantom{\nu}}^{\mu\phantom{\rho}\nu}$ can be expressed in terms of ${\sf g}^{\mu\nu}_{\phantom{\mu\nu},\rho}$ and $S^\rho_{\phantom{\rho}\mu\nu}$.
Consequently, we can identify $\textgoth{K}(g_{\mu\nu},g_{\mu\nu,\rho},S^\rho_{\phantom{\rho}\mu\nu},Q_{\mu\nu})$ with a Lagrangian density for matter $\mathcal{L}_M$ in the purely metric formulation of general relativity with torsion.\footnote{
This identification means that the purely affine theory based on the Lagrangian $\textgoth{L}$ and the purely metric theory based on the matter Lagrangian $\mathcal{L}_M$ equal to $\textgoth{K}$ have the same solutions to the corresponding field equations.
}
Similarly, the tensor $P_{\mu\nu}(\Gamma^{\,\,\rho}_{\mu\,\nu})$ in Eq.~(\ref{Ham2}) can be expressed as $P_{\mu\nu}(g_{\mu\nu},g_{\mu\nu,\rho},S^\rho_{\phantom{\rho}\mu\nu})$ which can be decomposed into the Riemannian Ricci tensor $\mathcal{R}_{\mu\nu}$ and terms with the torsion tensor~\cite{Scho}, yielding the standard form of the Einstein equations~\cite{FK3,nonsym,unif}.
The equivalence of purely affine gravity with general relativity, which is a metric theory, implies that the former is consistent with experimental tests of the weak equivalence principle~\cite{Wi}.

\section{F(R) Lagrangians}
\label{secFR}

The metric-affine Lagrangian density for the gravitational field $\mathcal{L}_g$ automatically turns out to be {\em linear} in the curvature tensor.
The purely metric Lagrangian density for the gravitational field also turns out to be linear in the curvature tensor: ${\sf L}_g=-\frac{1}{2\kappa}\mathcal{R}\sqrt{-g}$, since $P$ is a linear function of $\mathcal{R}=\mathcal{R}_{\mu\nu}g^{\mu\nu}$.
Thus metric-affine and metric Lagrangians for the gravitational field that are nonlinear with respect to curvature {\em cannot} be derived from a purely affine Lagrangian that depends on the connection and the contracted curvature tensors.
The reverse statement is also true: one cannot construct a dynamically equivalent, purely affine Lagrangian from a nonlinear (in curvature) metric-affine or metric Lagrangian.

To show this impossibility for both the metric-affine and metric pictures, it is sufficient to consider nonlinear metric-affine Lagrangians since the Legendre transformation with respect to the connection preserves the linearity of a Lagrangian with respect to the curvature, and to restrict the attention to $F(R)$ gravity models.
If we apply to a Hamiltonian density $\textgoth{H}(\Gamma^{\,\,\rho}_{\mu\,\nu},{\sf g}^{\mu\nu})$ the Legendre transformation with respect to ${\sf g}^{\mu\nu}$ (Refs.~\cite{Kij} and~\cite{FK3}) and use Eq.~(\ref{Ham1}):
\begin{equation}
\textgoth{L}=\textgoth{H}-\frac{\partial\textgoth{H}}{\partial {\sf g}^{\mu\nu}}{\sf g}^{\mu\nu}=\textgoth{H}-\frac{1}{2\kappa}{\sf g}^{\mu\nu}P_{\mu\nu},
\label{Leg3}
\end{equation}
we obtain a purely affine Lagrangian density $\textgoth{L}=\textgoth{L}(\Gamma^{\,\,\rho}_{\mu\,\nu},P_{\mu\nu})$.
However, if the metric-affine Lagrangian density for the gravitational field is a nonlinear function $F$ of the curvature scalar $R=P$:
\begin{equation}
\mathcal{L}_{g}=-\frac{1}{2\kappa}\sqrt{-g}F(R),
\label{fun1}
\end{equation}
then Eq.~(\ref{Ham1}) becomes
\begin{equation}
F'P_{\mu\nu}+\frac{1}{2}g_{\mu\nu}(F-RF')=2\kappa\frac{\partial\textgoth{H}}{\partial {\sf g}^{\mu\nu}},
\label{fun2}
\end{equation}
where $F'=\frac{dF(R)}{dR}$.
In this case, the variation of $\textgoth{L}$ is given by
\begin{equation}
\delta\textgoth{L}=-\frac{1}{2\kappa}\Pi_{\phantom{\mu}\rho\phantom{\nu}}^{\mu\phantom{\rho}\nu}\delta\Gamma^{\,\,\rho}_{\mu\,\nu}-\frac{1}{2\kappa}(F'-RF''){\sf g}^{\mu\nu}\delta P_{\mu\nu}-\frac{1}{2\kappa}\Bigl(\frac{1}{2}(F-RF'){\sf g}^{\mu\nu}+RF''P^{\mu\nu}\sqrt{-g}\Bigr)\delta g_{\mu\nu}.
\label{fun3}
\end{equation}
Consequently, the Lagrangian density $\textgoth{L}$ is purely affine if the expression that multiplies $\delta g_{\mu\nu}$ vanishes:
\begin{equation}
\frac{1}{2}(F-RF')g^{\mu\nu}+RF''P^{\mu\nu}=0.
\label{fun4}
\end{equation}
This condition is satisfied, without introducing additional constraints on the field, only if $F(R)=\alpha R$, where $\alpha=$ const, i.e. for the Einstein-Hilbert action.
In general, $F(R)$ gravity does not have a purely affine formulation.

A metric-affine Lagrangian density
\begin{equation}
\mathcal{L}=\mathcal{L}_g+\mathcal{L}_{MA}=\textgoth{H}-\frac{1}{2\kappa}\sqrt{-g}F(R)
\label{equ1}
\end{equation}
is dynamically equivalent to a scalar-tensor Lagrangian density
\begin{equation}
\mathcal{L}_{ST}=\textgoth{H}(\Gamma^{\,\,\rho}_{\mu\,\nu},g_{\mu\nu})-\frac{1}{2\kappa}\sqrt{-g}(F(\phi)+F^\cdot(\phi)(R-\phi)),
\label{equ2}
\end{equation}
where $\phi$ is a scalar field and $F^\cdot(\phi)=\frac{dF(\phi)}{d\phi}$.
This equivalence holds provided $F''(R)\neq0$ since the action variation with respect to $\phi$ yields $\phi=R$~\cite{st1,st2,st3,st4,EJ4}.
If we define a conformally transformed metric tensor:
\begin{equation}
\tilde{g}_{\mu\nu}=F^\cdot(\phi)g_{\mu\nu},
\label{equ3}
\end{equation}
then Eq.~(\ref{equ2}) becomes
\begin{equation}
\mathcal{L}_{ST}=\tilde{\textgoth{H}}(\Gamma^{\,\,\rho}_{\mu\,\nu},\tilde{g}_{\mu\nu},\phi)-\frac{1}{2\kappa}P_{\mu\nu}\tilde{g}^{\mu\nu},
\label{equ4}
\end{equation}
where
\begin{equation}
\tilde{\textgoth{H}}=\textgoth{H}-\frac{1}{2\kappa}(F^\cdot)^{-2}\sqrt{-\tilde{g}}(F-\phi F^\cdot).
\label{equ5}
\end{equation}
The tensor $\tilde{g}^{\mu\nu}$ is related to $\tilde{g}_{\mu\nu}$ by $\tilde{g}^{\mu\rho}\tilde{g}_{\nu\rho}=\delta^\mu_\nu$, $\tilde{g}$ denotes $\mbox{det}\tilde{g}_{\mu\nu}$ and $\tilde{{\sf g}}^{\mu\nu}=\sqrt{-\tilde{g}}\tilde{g}^{\mu\nu}$.
The action variation with respect to $\tilde{{\sf g}}^{\mu\nu}$ yields the Einstein equations:
\begin{equation}
P_{\mu\nu}=2\kappa\frac{\partial\tilde{\textgoth{H}}}{\partial\tilde{{\sf g}}^{\mu\nu}}.
\label{Ham4}
\end{equation}
The metric tensor $\tilde{g}_{\mu\nu}$ is referred to as the Einstein frame metric tensor, while $g_{\mu\nu}$ is said to define the Jordan frame~\cite{SL,EJ1,EJ2,EJ3,EJ4,EJ5,EJ6}.

If we apply to a Hamiltonian density $\tilde{\textgoth{H}}$ the Legendre transformation with respect to $\tilde{{\sf g}}^{\mu\nu}$ and use Eq.~(\ref{Ham4}):
\begin{equation}
\tilde{\textgoth{L}}=\tilde{\textgoth{H}}-\frac{\partial\tilde{\textgoth{H}}}{\partial\tilde{{\sf g}}^{\mu\nu}}\tilde{{\sf g}}^{\mu\nu}=\tilde{\textgoth{H}}-\frac{1}{2\kappa}\tilde{{\sf g}}^{\mu\nu}P_{\mu\nu},
\label{Leg4}
\end{equation}
we obtain
\begin{equation}
d\tilde{\textgoth{L}}=\frac{\partial\tilde{\textgoth{H}}}{\partial\Gamma^{\,\,\rho}_{\mu\,\nu}}d\Gamma^{\,\,\rho}_{\mu\,\nu}+\frac{\partial\tilde{\textgoth{H}}}{\partial\phi}d\phi-\frac{1}{2\kappa}\tilde{{\sf g}}^{\mu\nu}dP_{\mu\nu}.
\label{Leg5}
\end{equation}
Consequently, the Lagrangian density $\tilde{\textgoth{L}}$ is purely affine: $\tilde{\textgoth{L}}=\tilde{\textgoth{L}}(\Gamma^{\,\,\rho}_{\mu\,\nu},P_{\mu\nu},\phi)$ and equals $\mathcal{L}_{ST}$.
Therefore it is the Einstein frame metric tensor that plays the role of the generalized momentum in the Legendre transformation from the metric-affine (Hamiltonian) action to the purely affine (Lagrangian) action.

\section{Summary and Discussion}
\label{secSum}

In this paper we showed that metric-affine and metric $F(R)$ Lagrangians, with the exception of general relativity, cannot be related by a Legendre transformation to a purely affine Lagrangian that depends on the connection and the contracted curvature tensors.
This correspondence, however, is restored if we apply a conformal transformation of the metric tensor from the original Jordan frame, in which a Lagrangian is nonlinear in the Ricci scalar, to the Einstein frame, in which this Lagrangian becomes linear in the conformally transformed Ricci scalar and $F(R)$ gravity is turned into general relativity with a nonminimally coupled scalar field.
Therefore general relativity, among relativistic theories of gravitation, shows a peculiar behavior with respect to purely affine, metric-affine and purely metric variation: all three pictures are dynamically equivalent representations of the same theory like the Lagrange and Hamilton equations in classical mechanics.

In order to establish the physical frame, one should consider the frame in which physical quantities are measured and tested with respect to their theoretical predictions.
Although dynamical solutions of $F(R)$ models can be conformally mapped into the Einstein frame, it is not true for their physical properties since the Jordan-frame and Einstein-frame solutions corresponding to the same generic action can have different physical behaviors~\cite{equ1,equ2,equ3}.
We might conclude that the Einstein frame is {\em physical} since the affine/metric (or Lagrangian/Hamiltonian) equivalence works only in this frame.
However, it would be too much speculative deducing from this observation that nonlinear theories of gravity have no physical meaning and just represent a different description of Einstein's general relativity with some scalar fields, in which the scalar degree of freedom artificially enters the curvature part of the Lagrangian.
In the Einstein frame, matter is nonminimally coupled to the geometrical degrees of freedom and the scalar fields are of geometrical origin, inducing running coupling constants in the gravity sector and violations of the principle of equivalence which can serve as constraints on nonlinear theories of gravity.
The physically measured metric is determined from the coupling to matter, and the principle of equivalence can provide an operational definition of the metric tensor~\cite{Br}.
The question of which frame is physical should be ultimately answered by experiment or observation.

The result that $F(R)$ gravity does not allow an ambiguous formulation of its dynamical evolution with respect to different approaches can indicate, because of the higher number of allowed degrees of freedom, its intrisically richer structure.
On the other side, the fact that general relativity is the only relativistic theory of gravitation that allows such a dynamical equivalence, may indicate the uniqueness of this theory and could explain why general relativity cannot interpolate among phenomenological behaviors at different scales.


\begin{thebibliography}{99}
\bibitem{Ein} A. Einstein, {\em Sitzungsber. Preuss. Akad. Wiss.} ({\em Berlin}), 32 (1923).
\bibitem{Edd1} A. Einstein, {\em Sitzungsber. Preuss. Akad. Wiss.} ({\em Berlin}), 137 (1923).
\bibitem{Edd2} A. S. Eddington, {\em The Mathematical Theory of Relativity} (Cambridge Univ. Press, 1924).
\bibitem{Schr2} E. Schr\"{o}dinger, {\em Space-Time Structure} (Cambridge Univ. Press, 1950).
\bibitem{Kij} J. Kijowski, {\em Gen. Relativ. Gravit.} {\bf 9}, 857 (1978).
\bibitem{Cat} D. Catto, M. Francaviglia and J. Kijowski, {\em Bull. Acad. Polon. Sci.} ({\em Phys., Astron.}) {\bf 28}, 179 (1980).
\bibitem{Pal} A. Palatini, {\em Rend. Circ. Mat.} ({\em Palermo}) {\bf 43}, 203 (1919).
\bibitem{Lord} A. Einstein, {\it Sitzungsber. Preuss. Akad. Wiss.} ({\it Berlin}), 414 (1925).
\bibitem{FFR} M. Ferraris, M. Francaviglia and C. Reina, {\em Gen. Relativ. Gravit.} {\bf 14}, 243 (1982).
\bibitem{Hilb1} A. Einstein, {\em Sitzungsber. Preuss. Akad. Wiss.} ({\em Berlin}), 1111 (1916).
\bibitem{Hilb2} D. Hilbert, {\em K\"{o}nigl. Gesel. Wiss., G\"{o}ttingen Nachr., Math. Phys. Kl.}, 394 (1915).
\bibitem{Hilb3} H. A. Lorentz, {\em Konikl. Akad. Wetensch.} ({\em Amsterdam}) {\bf 25}, 468 (1916).
\bibitem{LL2} L. D. Landau and E. M. Lifshitz, {\em The Classical Theory of Fields} (Pergamon, 1975).
\bibitem{MTW} C. W. Misner, K. S. Thorne and J. A. Wheeler, {\em Gravitation} (Freeman, 1973).
\bibitem{SG} V. de Sabbata and M. Gasperini, {\em Introduction to Gravitation} (World Scientific, 1986).
\bibitem{FK3} M. Ferraris and J. Kijowski, {\em Gen. Relativ. Gravit.} {\bf 14}, 165 (1982).
\bibitem{FK4} M. Ferraris and J. Kijowski, {\em Rend. Sem. Mat. Univ. Polit. Torino} {\bf 41}, 169 (1983).
\bibitem{FK1} M. Ferraris and J. Kijowski, {\em Lett. Math. Phys.} {\bf 5}, 127 (1981).
\bibitem{KPT} J. Kijowski, B. Pawlik and W. M. Tulczyjew, {\em Bull. Acad. Polon. Sci.} ({\em Math., Phys., Astron.}) {\bf 27}, 163 (1979).
\bibitem{KM} J. Kijowski and G. Magli, {\em Class. Quantum Grav.} {\bf 15}, 3891 (1998).
\bibitem{KW} J. Kijowski and R. Werpachowski, {\em Rep. Math. Phys.} {\bf 59}, 1 (2007).
\bibitem{univ} A. Jakubiec and J. Kijowski, {\em J. Math. Phys.} {\bf 30}, 1073 (1989).
\bibitem{nonsym} A. Jakubiec and J. Kijowski, {\em J. Math. Phys.} {\bf 30}, 1077 (1989).
\bibitem{fR1} A. A. Starobinsky, {\em Phys. Lett. B} {\bf 91}, 99 (1980).
\bibitem{fR2} R. Kerner, {\em Gen. Relativ. Gravit.} {\bf 14}, 453 (1982).
\bibitem{fR3} J. D. Barrow and A. C. Ottewill, {\em J. Phys. A: Math. Gen.} {\bf 16}, 2757 (1983).
\bibitem{fR4} D. N. Vollick, {\em Phys. Rev. D} {\bf 68}, 063510 (2003).
\bibitem{fR5} X. Meng and P. Wang, {\em Class. Quantum Grav.} {\bf 20}, 4949 (2003).
\bibitem{fR6} S. Capozziello, S. Carloni and A. Troisi, {\em Rec. Res. Dev. Astron. Astrophys.} {\bf 1}, 25 (2003).
\bibitem{fR7} G. Allemandi, A. Borowiec and M. Francaviglia, {\em Phys. Rev. D} {\bf 70}, 043524 (2004).
\bibitem{fR8} S. M. Carroll, V. Duvvuri, M. Trodden and M. S. Turner, {\em Phys. Rev. D} {\bf 70}, 043528 (2004).
\bibitem{fR9} X. H. Meng and P. Wang, {\em Class. Quantum Grav.} {\bf 21}, 951 (2004).
\bibitem{fR10} G. Allemandi, A. Borowiec, M. Francaviglia and S. D. Odintsov, {\em Phys. Rev. D} {\bf 72}, 063505 (2005).  
\bibitem{fR11} S. Nojiri and S. D. Odintsov, {\em Phys. Rev. D} {\bf 74}, 086005 (2006).
\bibitem{fR12} V. Faraoni, {\em Phys. Rev. D} {\bf 74}, 023529 (2006).
\bibitem{Niko0} N. J. Pop\l awski, {\em Class. Quantum Grav.} {\bf 23}, 4819 (2006).
\bibitem{fR13} N. J. Pop\l awski, {\em Phys. Rev. D} {\bf 74}, 084032 (2006).
\bibitem{fR14} G. J. Olmo, {\em Phys. Rev. Lett.} {\bf 98}, 061101 (2007).
\bibitem{fR15} G. J. Olmo, {\em Phys. Rev. D} {\bf 75}, 023511 (2007).
\bibitem{fR16} S. Fay, R. Tavakol and S. Tsujikawa, {\em Phys. Rev. D} {\bf 75}, 063509 (2007).
\bibitem{fR17} L. Amendola, R. Gannouji, D. Polarski and S. Tsujikawa, {\em Phys. Rev. D} {\bf 75}, 083504 (2007).
\bibitem{fR18} K. Uddin, J. E. Lidsey and R. Tavakol, {\em Class. Quantum Grav.} {\bf 24}, 3951 (2007).
\bibitem{SL} T. P. Sotiriou and S. Liberati, {\em Ann. Phys.} ({\em N.Y.}) {\bf 322}, 935 (2007).
\bibitem{fR19} W. Hu and I. Sawicki, {\em Phys. Rev. D} {\bf 76}, 064004 (2007). 
\bibitem{fR20} C. G. Boehmer, L. Hollenstein and F. S. N. Lobo, {\em Phys. Rev. D} {\bf 76}, 084005 (2007). 
\bibitem{fR21} Y.-S. Song, H. Peiris and W. Hu, {\em Phys. Rev. D} {\bf 76}, 063517 (2007).
\bibitem{fR22} L. Amendola and S. Tsujikawa, {\it Phys. Lett. B} {\bf 660}, 125 (2008).
\bibitem{fR23} S. Carloni, A. Troisi and P. K. S. Dunsby, arXiv:0706.0452 [gr-qc].
\bibitem{fR24} M. Abdelwahab, S. Carloni and P. K. S. Dunsby, arXiv:0706.1375 [gr-qc].
\bibitem{fR25} S. Carloni, P. K. S. Dunsby and A. Troisi, {\it Phys. Rev. D} {\bf 77}, 024024 (2008).
\bibitem{fR26} S. Capozziello, R. Cianci, C. Stornaiolo and S. Vignolo, arXiv:0708.3038 [gr-qc].
\bibitem{fR27} T. P. Sotiriou, arXiv:0710.4438 [gr-qc].
\bibitem{EJ1} K.-I. Maeda, {\em Phys. Rev. D} {\bf 39}, 3159 (1989).
\bibitem{EJ2} G. Magnano and L. M. Soko\l owski, {\em Phys. Rev. D} {\bf 50}, 5039 (1994).
\bibitem{EJ3} V. Faraoni and S. Nadeau, {\em Phys. Rev. D} {\bf 75}, 023501 (2007).
\bibitem{Scho} J. A. Schouten, {\em Ricci-Calculus} (Springer-Verlag, 1954).
\bibitem{FK2} M. Ferraris and J. Kijowski, {\em Gen. Relativ. Gravit.} {\bf 14}, 37 (1982).
\bibitem{Niko1} N. J. Pop\l awski, arXiv:gr-qc/0701176.
\bibitem{Niko2} N. J. Pop\l awski, {\it Int. J. Mod. Phys. A} {\bf 23}, 567 (2008), arXiv:gr-qc/0702129.
\bibitem{Hehl} F. W. Hehl, P. von der Heyde, G. D. Kerlick and J. M. Nester, {\em Rev. Mod. Phys.} {\bf 48}, 393 (1976).
\bibitem{Smal} L. L. Smalley, {\em Phys. Lett. A} {\bf 61}, 436 (1977).
\bibitem{HK} F. W. Hehl and G. D. Kerlick, {\em Gen. Relativ. Gravit.} {\bf 9}, 691 (1978).
\bibitem{HLS} F. W. Hehl and E. A. Lord, L. L. Smalley, {\em Gen. Relativ. Gravit.} {\bf 13}, 1037 (1981).
\bibitem{Car} \'{E}. Cartan, {\em Ann. Ec. Norm. Sup.} {\bf 40}, 325 (1923).
\bibitem{unif} N. J. Pop\l awski, arXiv:0705.0351 [gr-qc].
\bibitem{Schr1} E. Schr\"{o}dinger, {\em Proc. R. Ir. Acad. A} {\bf 51}, 163 (1947).
\bibitem{LL1} L. D. Landau and E. M. Lifshitz, {\em Mechanics} (Pergamon, 1960).
\bibitem{PO} V. N. Ponomariev and Yu. N. Obukhov, {\em Gen. Relativ. Gravit.} {\bf 14}, 309 (1982).
\bibitem{Wi} C. M. Will, {\em Theory and Experiment in Gravitational Physics} (Cambridge Univ. Press, 1992).
\bibitem{st1} P. Teyssandier and P. Tourrenc, {\em J. Math. Phys.} {\bf 24}, 2793 (1983).
\bibitem{st2} B. Whitt, {\em Phys. Lett. B} {\bf 145}, 176 (1984).
\bibitem{st3} A. Jakubiec and J. Kijowski, {\em Phys. Rev. D} {\bf 37}, 1406 (1988).
\bibitem{st4} D. Wands, {\em Class. Quantum Grav.} {\bf 11}, 269 (1994).
\bibitem{EJ4} G. Magnano, arXiv:gr-qc/9511027.
\bibitem{EJ5} V. Faraoni and E. Gunzig, {\em Int. J. Theor. Phys. D} {\bf 38}, 217 (1999).
\bibitem{EJ6} \'{E}. \'{E}. Flanagan, {\em Class. Quantum Grav.} {\bf 21}, 3817 (2004).
\bibitem{equ1} V. Faraoni, {\em Phys. Rev. D} {\bf 62}, 023504 (2000).
\bibitem{equ2} S. Capozziello, S. Nojiri, S. D. Odintsov and A. Troisi, {\em Phys. Lett. B} {\bf 639}, 135 (2006).
\bibitem{equ3} G.Allemandi, M. Capone, S. Capozziello and M. Francaviglia, {\em Gen. Relativ. Gravit.} {\bf 38}, 33 (2006).
\bibitem{Br} C. H. Brans, {\em Class. Quantum Grav.} {\bf 5}, L197 (1988).
\end{thebibliography}
\end{document}